\documentclass{article}

\usepackage{amssymb,amsmath,xcolor,graphicx,xspace,colortbl, rotating} 
\usepackage{boxedminipage}  
\usepackage{graphics}  
\usepackage{ragged2e}  
\usepackage{xcolor}  
\graphicspath{{numereval_graphics/}{numereval_tcache/}{numereval_gcache/}}
\DeclareGraphicsExtensions{.pdf,.svg,.eps,.ps,.png,.jpg,.jpeg}

\begin{document}

\title{Mean velocity equation for fluctuating flow:\newline Numerical evaluation for small Reynolds numbers }
\author{
              Juergen Piest}
\maketitle

\begin{abstract}The mean velocity equation for fluctuating flow derived in an earlier paper of the author has been applied to circular jet flow for
low Reynolds numbers. Numerical solutions have been obtained from an iteration starting with the corresponding solution of the Navier-Stokes equation. Deviations
become distinct for $\ensuremath{\operatorname*{Re}} >30$
\end{abstract}

\section{
              Introduction}
This paper
deals with the closure problem of turbulence theory. The problem is encountered, in its simplest form, when one attempts to derive an equation for the mean
velocitiy field in turbulent flow. In an earlier paper \cite{pi2013}, the
author suggested to employ classical statistical mechanics of the particle cloud picture of the fluid in order to derive the equation. The reason is the
probabilistic structure of this picture which is suited to an investigation of an expectation value equation. An important feature is that in statistical
mechanics the hydrodynamic velocity $\boldsymbol{u}$ itself is an expectation. The consequence is that, again in the frame of statistical mechanics, the definitions of hydrodynamic velocity
and its expectation coincide. A dynamic equation for this quantity is derived, by applying projection operator technique, in the work of Zwanzig an
Mori. In the monograph of H. Grabert \cite{gra}, an unified form of the
derivation is presented; the equation is a specialization of the Generalized Transport Equation.

While
the advective part of the equation has the well known second-order form, the dissipative part in general is a non-linear functional of the velocity field.
Grabert shows that when one linearizes the dissipative term, one obtains the Stokes form; thus, the equation then coincides with the Navier-Stokes equation.
It follows that the difference between the equations of~~$\boldsymbol{u}$ and its expectation is, in the frame of statistical mechanics, the appearance of the nonlinear
part of the dissipation term.

The dissipative term of the equation has the form of a convolution integral, the kernel of which is a local equilibrium correlation. Presently, there is
no theory for the calculation of this quantity; as a consequence, for the time being, the equation cannot be used in general form. The author expanded the
dissipation term to the lowest non-linear order in the velocity (which is third order); the kernel now appears to be a total equilibrium correlation, which
can be calculated. There is a Reynolds number range where the nonlinear dissipation term is small against the linear one. This is the range of the validity
of the Navier-Stokes equation. The equation with the nonlinearly approximated dissipation term should be applied for Reynolds numbers which do not excede
too far this range.

There is another restrictive feature (already mentioned in \cite{pi2013})
of the theory in its present form: The statistical mechanics formalism is, as usual, for a fluid with no boundaries; the same is true for the correlation
function calculations. In Navier-Stokes calculations, boundaries are accounted for by boundary conditions, but do not change the form of the equation. Here
the situation is different: The existence of a boundary should influence the formulas for correlation functions, and via the dissipation term alter the form
of the equation itself. The author has attempted to calculate the stationary flow in a tube with the formulas in their present form, and obtained the result
that the third-order friction term is generally zero for this configuration - apparantly a non-physical feature. The possibility to calculate correlation
functions in the presence of boundaries is one of the major necessities for making this theory applicable.

The final form of the equation
is(\cite{pi2013} (2.13)): \begin{equation}\rho (\frac{ \partial \boldsymbol{u}}{ \partial t} +\boldsymbol{u} \cdot  \nabla \boldsymbol{u}) = - \nabla P +\nu \rho  \nabla ^{2}\boldsymbol{u} +\boldsymbol{D} \label{1.1}
\end{equation}

where\ $\rho $ is the density,~$P$ the pressure,~$\nu $ the kinematic viscosity, and~$\boldsymbol{D}$, the nonlinear part of the friction force,\ is given by \cite{pi2013}
(10.10-10.15). In the following sections, this equation will be called the momentum equation. Together with the continuity equation, it describes incompressible
flow with constant density and temperature. This is the starting point for the present paper. - The theoretical formulas have been tested for the circular
jet as a representative flow, since it is thought that in this case boundary effects may be relatively small. Jet flow for small Reynolds numbers is not
turbulent, but still represents a fluctuating flow situation; so it is a candidate for application of equation (\ref{1.1}).
- For the present investigation, the author used a personal computer algebra system (Mathematica). As will be seen in the next section, the momentum equation
is actually an integro-differential equation. Presently a numerical solution seems not feasable; especially, there is no command in Mathematica which will
solve the equation automatically. - In this situation, it seems possible to use an existing empirical formula for the jet velocity field as a test formula
in order to investigate the various terms of the momentum equation, and perhaps find that the test formula is a solution. There are essentially the formulas
by Schlichting \cite{sch}\ and
by Schneider \cite{schn85}. The Schlichting formulas are valid, with suitable
coefficients, for laminar as well as for turbulent flow. For laminar flow, they are a solution of the Navier-Stokes equation under the condition of constant
momentum flux. Schlichting uses a point source at the orifice of the jet; this can be altered by introducing an ''effective point source'' as has been installed
by Andrade et al. \cite{at}\ in
order to secure convergence of the integrals in the momentum equation. But there is a severe impedimant: With the Schlichting formulas, the integrals definitely
diverge in the far region of the field. Thus, the present theoretical formulas are in contradiction with the Schlichting model. - Schneiders formulas are
for laminar flow, i. e. for rather small Reynolds numbers. They are the result of a theoretical investigation which leads to a slowly varying momentum flux.
The formulas show good agreement with the experimental results of Zauner \cite{z85}
which confirm that for these Reynolds numbers the round jet has essentially a finite length and is part of a recirculation pattern. With these formulas,
the integrals converge in the far field. But here another problem arose: The Schneider model again uses a point source at the orifice. The author did not
succeed in substituting this by an effective point source in such a way that the velocity field behaves reasonably in other parts of the flow space.

In order to account for the lack of a velocity field formula which can be used here, laminar jet flow fields for a set of Reynolds numbers have
been numerically calculated. These are stationary solutions of the Navier-Stokes equation. This is reported in sec. 3 of this paper. Calculations of jet
flow have already been performed 1985 by W. Schneider and co-workers \cite{bsz85};
many details, especially concerning boundary conditions, have been adopted from this paper. Some properties of the resulting flow fields are discussed and
compared with results of other publications mentioned before. In sec. 4, using these results, the nonlinear friction term is numerically calculated as a
function of axial and radial coordinates. The computation is impeded by the large scatter of the data, the reason for which in not well understood presently.
In sec. 5 an iterative procedure for solving the momentum equation is formulated, which enables an approximate numerical solution; the differences to the
solution of the Navier-Stokes equation are considered. Unfortunately, in this Reynolds number range, there are no experimental investigations where the
velocity field is measured. Some qualitative aspects mentioned in the paper of A. J. Reynolds \cite{re62}\ are
discussed.~

\section{Evaluation of the nonlinear dissipation term}
The formulas
for the third order dissipation term~are obtained as:

\begin{equation}D_{a}(x ,t) \prime  =\frac{\rho }{\nu }\int dx^{ \prime }dx^{ \prime  \prime }dx^{ \prime  \prime  \prime }\sum _{\mu  =1}^{2}M_{\mu }(\boldsymbol{x} -\boldsymbol{x}^{ \prime } ,\boldsymbol{x} -\boldsymbol{x}^{ \prime  \prime } ,\boldsymbol{x} -\boldsymbol{x}^{ \prime  \prime  \prime })\thinspace (Q_{\mu })_{a}(\boldsymbol{x}^{ \prime } ,\boldsymbol{x}^{ \prime  \prime } ,\boldsymbol{x}^{ \prime  \prime  \prime }) \label{2.1a}
\end{equation}

\begin{equation}M_{1}(\boldsymbol{x}^{ \prime } ,\boldsymbol{x}^{ \prime  \prime } ,\boldsymbol{x}^{ \prime  \prime  \prime }) =\frac{1}{2^{4}\pi ^{3}x^{ \prime }x^{ \prime  \prime }}\arctan (\genfrac{(}{)}{}{}{x^{ \prime }}{x^{ \prime  \prime }})\delta (\boldsymbol{x}^{ \prime } -\boldsymbol{x}^{ \prime  \prime  \prime }) \label{2.2}
\end{equation}

\begin{equation}M_{2}(\boldsymbol{x}^{ \prime } ,\boldsymbol{x}^{ \prime  \prime } ,\boldsymbol{x}^{ \prime  \prime  \prime }) =\frac{1}{2^{8}\pi ^{4}x^{ \prime  \prime }xt}\arctan (\genfrac{(}{)}{}{}{xs}{xr}) \label{2.3}
\end{equation}

\begin{equation}\boldsymbol{x}\boldsymbol{s} =\frac{1}{2}(\boldsymbol{x}^{ \prime } -\boldsymbol{x}^{ \prime  \prime  \prime }) \label{2.4}
\end{equation}

\begin{equation}\boldsymbol{x}\boldsymbol{t} =\frac{1}{2}(\boldsymbol{x}^{ \prime } +\boldsymbol{x}^{ \prime  \prime  \prime }) \label{2.5}
\end{equation}

\begin{equation}xr =x^{ \prime  \prime } +xt \label{2.6}
\end{equation}

\begin{multline}(Q_{1})_{a} =(s_{3})_{aceg}(s_{3})_{bdfh}( \nabla _{g}^{ \prime } + \nabla _{g}^{ \prime  \prime } + \nabla _{g}^{ \prime  \prime }) \nabla _{h}^{ \prime }\varepsilon _{ef}( \nabla ^{ \prime } + \nabla ^{ \prime  \prime  \prime }) \times  \\
\times u_{b}(\boldsymbol{x}^{ \prime } ,t)u_{c}(\boldsymbol{x}^{ \prime  \prime } ,t)u_{d}(\boldsymbol{x}^{ \prime  \prime  \prime } ,t) \label{2.7}\end{multline}

\begin{multline}(Q_{2})_{a} =(s_{3})_{aceg}(s_{3})_{fbdh}( \nabla _{g}^{ \prime } + \nabla _{g}^{ \prime  \prime } + \nabla _{g}^{ \prime  \prime  \prime })( \nabla _{h}^{ \prime } + \nabla _{h}^{ \prime  \prime  \prime }) \nabla ^{ \prime  \prime  \prime 2}\varepsilon _{ef}( \nabla ^{ \prime } + \nabla ^{ \prime  \prime  \prime }) \times  \\
\times u_{b}(\boldsymbol{x}^{ \prime } ,t)u_{c}(\boldsymbol{x}^{ \prime  \prime } ,t)u_{d}(\boldsymbol{x}^{ \prime  \prime  \prime } ,t)\end{multline}

\begin{equation}\varepsilon _{ab}( \nabla ) = \nabla ^{2}\delta _{ab} - \nabla _{a} \nabla _{b} \label{2.9}
\end{equation}

\begin{equation}(s_{3})_{abcd} =\delta _{ab}\delta _{cd} +\delta _{ac}\delta _{bd} -\lambda \delta _{ad}\delta _{bc} \label{2.10}
\end{equation}

Latin indices run from 1 to 3. A bold symbol indicates a vector; the corresponding non-bold symbol
its magnitude.~$\delta $\ is the Dirac function,~$\delta _{ab}$\ the Kronecker symbol.~$\lambda $\ is a physical constant of the fluid given by \cite{pi2013}\ (9.6);\ it
can be reduced to:

\begin{equation}\lambda  =\frac{\alpha }{\gamma \thinspace c_{V}} \label{2.11}
\end{equation}

Here,~$\alpha $\ is the thermal expansion coefficient,~$\gamma $\ the isothermal compressibility, $c_{V}$\ the specific heat for constant volume. For standard conditions, we have~$\lambda  =0.40$\ for air and~$\lambda  =0.091$\ for water. By this quantity, the results depend to a certain extent on the physical
properties of the fluid. The calculations presented in this paper are for a hypothetical fluid with $\lambda  =0$. - As is seen, (\ref{2.1a}) is a nonlinear convolution integral. It consists of two main terms
with kernel functions~$M_{1}$,~$M_{2}$ and velocity functions~$\boldsymbol{Q}_{1}$,~$\boldsymbol{Q}_{2}$. The velocity functions have been expanded; they consist of numerous terms with velocity gradients up to~$4^{th}$ and~$6^{th}$ order, respectively.

Generally, formulas (\ref{1.1}), (\ref{2.1a})
are 3 nonlinear partial integro-differential equations of the Fredholm type; together with the continuity equation, they form a system of 4 equations for
the three components of the velocity and the pressure. - For the application in mind, all formulas have been tranformed to cylinder coordinates and specified
to cylinder symmetric conditions. For most circular jet investigations, boundary layer conditions are considered to be valid; that is, radial components
are small compared to axial components, and axial gradients are small to radial ones. In a jet, this will be true for regions not too far from the nozzle.
In the present investigaton, the space integrals appearing in (\ref{2.1a}) extend over the whole flow area; but major contributions
to the integral will be from the region in the neighbourhood of the axis and not too far from the nozzle. Thus, boundary layer conditions may still be a
reasonable approximation. The formulas then considerably simplify.

An orifice of diameter~$d$\ exhibiting a parabolic velocity distribution with center velocity $u_{0}$\ is introduced. In (\ref{1.1}), all quantities are made
dimensionless by using as parameters~$d$ and~$u_{0}$. Then the third order term, instead of $\frac{\rho }{\nu }$, is proportional to the Reynolds number $\ensuremath{\operatorname*{Re}} =\frac{u_{0}d}{\nu }$. - It will be explained in sec. 4 that for detailed investigations we will only need the axial components of the two terms of (\ref{2.1a}),
now called~$D_{1}$,~$D_{2}$ which are now functions of the axial and radial coordinates~$x$, $r$.~ $D_{1}$\ simplifies from the Dirac function in (\ref{2.2})
and finally reduces to the form:

\begin{multline}D1(x ,r) =\frac{\ensuremath{\operatorname*{Re}}}{16\pi ^{3}}\int \limits _{0}^{\infty }dx^{ \prime }\int _{0}^{\infty }dr^{ \prime }\int _{0}^{\infty }dx^{ \prime  \prime }\int _{0}^{\infty }dr^{ \prime  \prime }\int _{0}^{2\pi }d\varphi ^{ \prime }\int _{0}^{2\pi }d\varphi ^{ \prime  \prime } \times  \\
\times \frac{r^{ \prime }r^{ \prime  \prime }}{z^{ \prime }z^{ \prime  \prime }}\arctan (\genfrac{(}{)}{}{}{z^{ \prime }}{z^{ \prime  \prime }})Q1(x^{ \prime } ,r^{ \prime } ,x^{ \prime  \prime } ,r^{ \prime  \prime }) \label{2.12}\end{multline}

\begin{equation}z^{ \prime } =\sqrt{(x -x^{ \prime })^{2} +r^{2} +r^{ \prime 2} -2rr^{ \prime }\cos (\varphi ^{ \prime })} \label{2.13}
\end{equation}

\begin{equation}z^{ \prime  \prime } =\sqrt{(x -x^{ \prime  \prime })^{2} +r^{2} +r^{ \prime  \prime 2} -2rr^{ \prime  \prime }\cos (\varphi ^{ \prime  \prime })} \label{2.14}
\end{equation}

The evaluated function~$Q1$ has been denoted with the same symbol as in (\ref{2.6}); it will be further specified in sec.
4. The formula for~$D2$\ has been elaborated in the same way but will not be given here.

\section{Numerical calculation of laminar jet flow}
As is
stated in the introduction, the author performed numerical calculations of the velocity field of circular jet flow which are stationary solutions of the
Navier-Stokes equation. The computations of \ W. Schneider and co-workers \cite{bzs85}
are for momentum Reynolds numbers $\tilde{R} =3.5$, $5.5$:

\begin{equation}Rm =\frac{\sqrt{m}}{\nu }
\end{equation}

$m$~is the momentum flux at the orifice divided by 2$\pi $. The relation to~$\ensuremath{\operatorname*{Re}}$\ is:

\begin{equation}\ensuremath{\operatorname*{Re}} =Rm\sqrt{24} \label{3.2}
\end{equation}

So, the numbers from above correspond to $\ensuremath{\operatorname*{Re}} =17$, $27$. For the present purpose, it is necessary to repeat the calculations for some other Reynolds numbers. - The author intended
to perform the calculations with the aid of the NDSolve command of Mathematica. There is now a choice between the so-called line method which uses a rectangular
net with constant mesh size, and the finite element method. But in NDSolve, the finite element method is presently restricted to linear equations of at
most second order. Thus it was necessary to use the line method. This method cannot proceed equations of the elliptic type; so it was not possible to use
the stationary Navier-Stokes equation. It was decided to solve the instationary equation up to a time where further variations with time could be neglected,
so that the final solution could be declared to be the stationary one.

For circular symmetric flow, in cylinder coordinates there
exists the Stokes stream function:

\begin{equation}\begin{array}{c}u =\frac{1}{r} \partial _{r}\psi  \\
v = -\frac{1}{r} \partial _{x}\psi \end{array} \label{3.3}
\end{equation}

We switch to non-dimensional quantities, as is already done in the nonlinear dissipation term.
Especially, the time is made non-dimensional with~$\frac{d}{u_{0}}$. To the Navier-Stokes equation, there corresponds the stream function equation (Milne-Thomson,  \cite{mt},
p. 647):

\begin{equation} \partial _{t}\zeta  -\frac{ \partial (\psi  ,\frac{1}{r}\zeta )}{ \partial (x ,r)} -\frac{1}{\ensuremath{\operatorname*{Re}}}\frac{1}{r}E^{2}(r\zeta ) =0 \label{3.4}
\end{equation}

\begin{equation}\zeta  = -\frac{1}{r}E^{2}\psi  \label{3.5}
\end{equation}

\begin{equation}E^{2}f = \partial _{x ,x}f +r \partial _{r}(\frac{1}{r} \partial _{r}f) \label{3.6}
\end{equation}

with~$\zeta $ being the third component of the curl:

\begin{equation}\zeta  =( \nabla  \times \boldsymbol{u})_{3} \label{3.7}
\end{equation}
              The different sign of the second term of (\ref{3.4}) in \cite{mt}\ is
due to the different sign in the definition of~$\psi $ by Milne-Thomson. Corresponding to the momentum equation, we have the stream
function equation:

\begin{equation} \partial _{t}\zeta  -\frac{ \partial (\psi  ,\frac{1}{r}\zeta )}{ \partial (x ,r)} -\frac{1}{\ensuremath{\operatorname*{Re}}}\frac{1}{r}E^{2}(r\zeta ) =\ensuremath{\operatorname*{Re}}\Delta  \label{3.8}
\end{equation}\begin{equation}\Delta  =( \nabla  \times \boldsymbol{D})_{3} \label{3.9}
\end{equation}

Instead of determining~$u$,~$v$ directly, the stream functionhas been computed by solving equation (\ref{3.4}). The calculation
is performed for a square of dimensionless length $a$. The boundary conditions have been copied from \cite{bsz85}.
The orifice has the width $\frac{1}{2}$; thus for the wall we have the condition:

\begin{equation}u(0 ,r) =\left \{\begin{array}{c}1 -4r^{2}\text{~~~ ,}\text{}\text{~~}0 \leq r \leq 0.5 \\
0\text{~~~~~~~ ,}\text{}\text{
~}0.5 <r\end{array}\right . \label{3.10}
\end{equation}

On the jet axis, the conditions read:

\begin{equation}\begin{array}{c} \partial _{r}u(x ,r)\vert _{r =0} =0 \\
v(x ,0) =0\end{array} \label{3.11}
\end{equation}

For the stream function $\psi (t ,x ,r)$, the calculation starts from rest:

\begin{equation}\psi (0 ,x ,r) =0 \label{3.12}
\end{equation}

Conditions (\ref{3.10}), (\ref{3.11})
are transformed for~$\psi $:

\begin{equation}\psi (t ,0 ,r) =\left \{\begin{array}{c}\frac{r^{2}}{2} -r~ \text{}
\text{~ ,~}\text{}\text{}0 \leq r \leq 0.5 \\
\frac{1}{16}\text{~ ,~}0.5 <r\text{
}\end{array}\right .\text{} \label{3.13}
\end{equation}

\begin{equation}\begin{array}{c}(r \partial _{r ,r}\psi (t ,x ,r) - \partial _{r}\psi (t,x,r))\vert _{r =0} =0 \\
\psi (t ,x ,0) =0\end{array} \label{3.14}
\end{equation}

Unfortunately, it is necessary to prescribe conditions for the outer boundaries of the square
also. This is a consequence of the calculation being limited to a finite area while the true motion extends over the total positive $x$, $r$-space. Some estimate for the stream function has to be found for the outer boundaries, for a suitable chosen square width $a$, so that the solution can be trusted for, say $x ,r <\frac{a}{2}$. Schneider and co-authors used the analytical stationary solution for the linearized Navier-Stokes equation (creeping flow), which
they took from \cite{hb73}, p. 153; in polar coordinates:

\begin{equation}\psi (r ,\theta ) =k(1 -\cos (\theta )^{3}) \label{3.15}
\end{equation}
              The constant~$k$ is determined from the value of~$\psi $\ for~$\theta  =\frac{\pi }{2}$\ from (\ref{3.13}):~$k =\frac{1}{16}$ . - From this, the conditions for the outer boundaries are inferred:

\begin{equation}\psi (t ,a ,r) =\frac{1}{16}(1 -\frac{a^{3}}{(a^{2} +r^{2})^{\frac{3}{2}}}) \label{3.16}
\end{equation}

\begin{equation} \partial _{x}\psi (t ,x ,r)\vert _{x =a} = -\frac{3r^{2}a^{2}}{16(a^{2} +r^{2})^{\frac{5}{2}}} \label{3.17}
\end{equation}

\begin{equation}\psi (t ,x ,a) =\frac{1}{16}(1 -\frac{x^{3}}{(x^{2} +a^{2})^{\frac{3}{2}}}) \label{3.18}
\end{equation}

\begin{equation} \partial _{r}\psi (t ,x ,r)\vert _{r =a} =\frac{3ax^{3}}{16(x^{2} +a^{2})^{\frac{5}{2}}} \label{3.19}
\end{equation}

The boundary conditions have somewhat adapted for $t <1$ in order to conform them to the initial condition (\ref{3.12}). - Calculations were performed
for a set of Reynolds numbers. Values for $a$ started with the values chosen in \cite{bsz85};\ a
suitable value for the end time~$t_{e}$\ of the instationary calculation was found by determining the coordinates of the center of
the toroidal eddy for a set of end times and observe for which $t$-values they become steady.

Since from now we only consider results
obtained for the final time, we cease writing the time as an argument. - A disturbing feature of the results was that the value~$u(0 ,0) =1$ prescribed by the boundary conditions is not returned by the calculations; for most Reynolds numbers, it was considerably smaller.
It became apparent that the computed value depends strongly on the width~~$a$ chosen for the calculation, but nearly not on the Reynolds number. Smaller $a$~led to a larger center velocity. There is a optimal value~$a_{0} =56.4$\ where the boundary condition is met exactly, again nearly independent of the Reynolds
number. - In a second run, the procedure was somewhat altered: After the calculation with individually chosen~$a$, a second run with~$a_{0}$\ was performed, where for the outer boundaries the results from the first run were taken;
finally, the $\psi $\ field was composed from the two results. - As an example, the result for~$\ensuremath{\operatorname*{Re}} =27$\ is shown, for comparison with the result in  \cite{bzs85};
the data are~$a =200$~\ and~$t_{e} =100000$ . Fig. 1 is the contour plot of the stream function, which shows the stream lines of the flow:

\begin{center}\fcolorbox[HTML]{FFFFFF}{FFFFFF}{\includegraphics[]{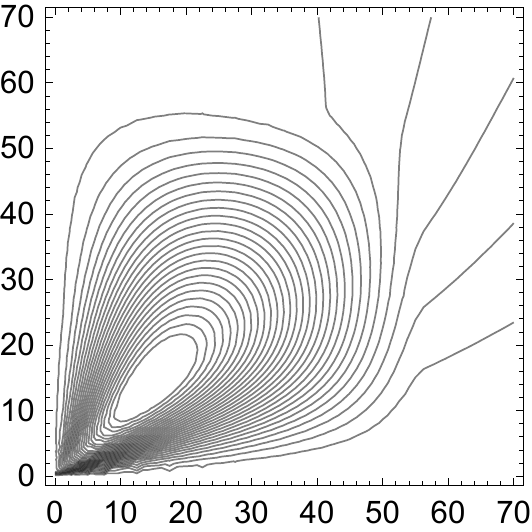}}\end{center}

\begin{center}Fig. 1: Contour plot of the stream function $\psi (x ,r)$,~ $\ensuremath{\operatorname*{Re}} =27$
\end{center}\par
\bigskip

The similarity to \cite{bsz85}\ Fig.
5 is visible; but the distance of the maximum from the orifice,~$r_{m}$, is somewhat larger here. It is seen how the streamlines are composed at the outer boundaries of the inner region. The irregularities
of the stream lines in the neighborhood of the orifice have been produced by the program for preparing the contour plot; if you ask for a plot for, say,~$x ,r <30$\ , the lines are smooth. - It is interesting to compare the values~$r_{m}$\ obtained for several~$\ensuremath{\operatorname*{Re}}$\ with a theoretical formula given by W. Schneider \cite{schn85}
              
              :

\begin{equation}r_{m} =b \exp  (\frac{\ensuremath{\operatorname*{Re}}^{2}}{192 C^{2}})
\end{equation}

~Fig. 2 shows calculated values
$r_{m}$\ points for a set of Reynolds numbers:

\begin{center}\fcolorbox[HTML]{FFFFFF}{FFFFFF}{\includegraphics[]{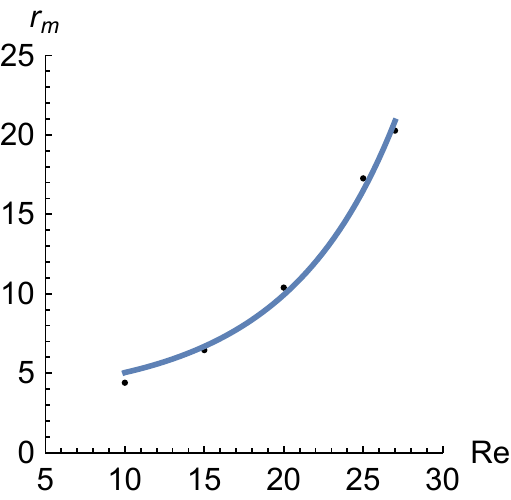}}\end{center}

\begin{center}Fig. 2: Calculated values~$r_{m}$\ against $R e$\end{center}\par

\begin{center}
\end{center}\par

The constants of best fit for the calculated values here are $b =4.02$,~$C^{2} =2.31$.~$b$\ remains undetermined in Schneiders theory,~$C^{2}$ is a value between analytical and experimental valus in \cite{bsz85}\ Fig.
6. Thus, for this Reynolds number range, the calculated valus fit reasonably in Schneiders theoretical work. For larger Reynolds numbers, the method of
composing the stream function field cannot be used any more: In the inner region, the eddy center starts to 'feel' the artificial right boundary and gets
distorted. Then it is presumably necessary to employ a solver which permits variable net width.

In experimental jet investigations,
often the axial velocity at the jet axis is considered. Fig. 3 shows the behavior due to several authors, in nondimensional variables, for $\ensuremath{\operatorname*{Re}} =27$:

\begin{center}\fcolorbox[HTML]{FFFFFF}{FFFFFF}{\includegraphics[]{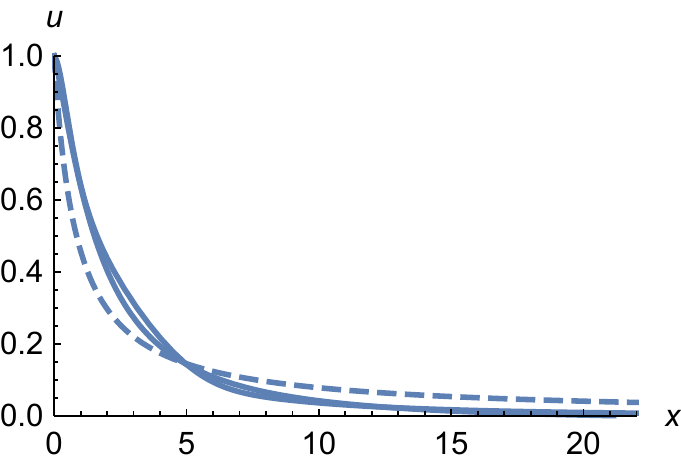}}\end{center}

\begin{center}    Fig. 3: Velocity at the jet axis. Re=27
              
              \end{center}\par

\begin{center}
\end{center}\par
The dashed line is from Schlichtings formula, with a suitably introduced effective point source in order to account for an orifice
with finite width. The full lines are from Schneiders theoretical investigations which interpret the experimental results in \cite{z85},
and the result of the present calculation. It is not possible to evaluate it directly from the stream function data at~$r =0$, because of the factor $\frac{1}{r}$\ in (\ref{3.3}). From visual interpretation, it became
apparent that the values $u(x ,r_{a})$\ with $r_{a} =0.0001$\ can be used as the axis velocity. The full lines are only marginally different.

\section{Numerical calulation of third order term using the results for laminar flow}
We are
now in the position to perform a calculation of the third order term using the numerical results of the preceding section. The first attempt has been done
with the results for~$\ensuremath{\operatorname*{Re}} =20$. From the velocity function~$Q1$ in (\ref{2.12}), a single example term~$Q1x$\ has been chosen:

\begin{equation}Q1x =u(x^{ \prime  \prime } ,r^{ \prime  \prime }) \partial _{x}u(x^{ \prime } ,r^{ \prime }) \partial _{x ,r ,r}u(x^{ \prime } ,r^{ \prime }) \label{4.1}
\end{equation}

For this preliminary calculation, it was sufficient to use only the outer field. The gradients
appearing in (\ref{4.1}) were computed as Mathematica InterpolatingFunctions, and then it was attempted to calculate the
corresponding partial term of $D1$\ (\ref{2.12}) with the aid of the Mathematica command
NIntegrate. The result clearly showed that the integral diverges somewhere in the integration region. - It was presumed that because of the $\frac{1}{r}$\ factors in (\ref{3.3}) some numerical errors have blown
up in the neighbourhood of the jet axis. As an example, the laminar friction term $la$\ has been calculated for several values of~$x$ as a function of $r$. Fig. 4 shows two curves for~$x =5.51$ and $x =5.52$:

\begin{center}\fcolorbox[HTML]{FFFFFF}{FFFFFF}{\includegraphics[]{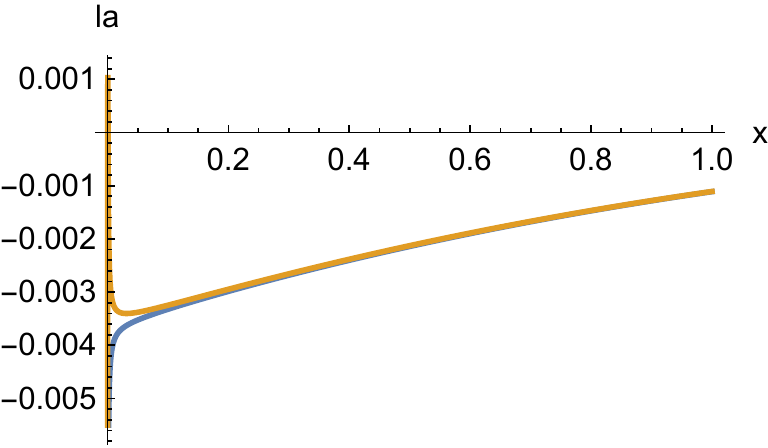}}\end{center}

\begin{center}    Fig. 4: Laminar friction force for$x =5.51$,~ $x =5 ,52$ as function of$r$; $\ensuremath{\operatorname*{Re}} =20$\end{center}\par
It is seen that the curves exhibit irregular singularities at the jet axis, in this case for $r <0.02$. For other $x$-values, it was necessary to look for the behavior very close to the axis. The final prescriptions for smoothing the velocity
components in the neighbourhood of the axis were chosen:

\begin{equation}u_{s}(x ,r) =\left \{\begin{array}{c}u(x ,r)\text{~ ,~}r \geq 0.0001\text{
} \\
u(x ,0.0001)\text{~ ,~}r <0.0001\text{
}\end{array}\right . \label{4.2}
\end{equation}
              \begin{equation}v_{s}(x ,r) =\left \{\begin{array}{c}v(x ,r)\text{~ ,~}r \geq 0.0001\text{
} \\
\text{0~ ,~}r <0.0001\text{
}\end{array}\right . \label{4.3}
\end{equation}

The smoothed components are called~$u$,~$v$ again. After this manipulations, the computed integrals converged. - As a first application, the first part~$D1$\ of the third order term (formula (\ref{2.12}))\ has
been calculated. The velocity function originally consisted of two parts:

\begin{equation}Q1 =q0 +q1\cos (\varphi ^{ \prime } -\varphi ^{ \prime  \prime }) \label{4.4}
\end{equation}

Both parts have been computed separately, for $x =30$, $r =5$. The second term amounted to about~$0.5 \%$ of the first one. For further calculations, this second part has been neglected. This has already been allowed
for in the form of (\ref{2.12}). For~$q0$\ a lengthy formula is obtained which is presented in the appendix. - Then, the two
parts of the third order term~$D1$,~$D2$\ mentioned in sec. 2 were calculated separately, on the jet axis for a set of~$x$-values, for~$\ensuremath{\operatorname*{Re}} =35$ as an example. Larger values for~$D1$\ are of the order~$10^{ -3}$\ to~$10^{ -4}$, while $D2$-values generally are of $10^{ -7}$. Thus,~$D2$\ will be neglected in the numerical calculations.

In order to
solve the stream function equation for the mean velocity field (\ref{3.8}), we need to calculate the quantity~$\Delta $ (\ref{3.9}):

\begin{equation}\Delta  = \partial _{x}D_{2} - \partial _{r}D_{1} \label{4.5}
\end{equation}

In boundary layer approximation, the first term will be two orders lower than the second, so that
we approximate:

\begin{equation}\Delta  = - \partial _{r}D_{1} \label{4.6}
\end{equation}

As has been mentioned, the axial component~$D_{1}$ is approximated by~$D1$; so the final formula is:

\begin{equation}\Delta  = - \partial _{r}D1 \label{4.7}
\end{equation}

where~$D1$\ is given by (\ref{2.12}).~$\Delta $\ has been calculated by differentiating analytically the kernel function in
(\ref{2.12}), obtaining:~

\begin{equation}\Delta (x ,r) = -\frac{\ensuremath{\operatorname*{Re}}}{16\pi ^{3}}\int _{0}^{\infty }dx^{ \prime }\int _{0}^{\infty }dr^{ \prime }\int _{0}^{\infty }dx^{ \prime  \prime }\int _{0}^{\infty }dr^{ \prime  \prime }\int _{0}^{2\pi }d\varphi ^{ \prime }\int _{0}^{2\pi }d\varphi ^{ \prime  \prime }DM1\thinspace q0 \label{4.8}
\end{equation}

\begin{multline}DM1 =\frac{r^{ \prime }r^{ \prime  \prime }}{z^{ \prime 2}z^{ \prime  \prime 2}}( -\arctan (\genfrac{(}{)}{}{}{z^{ \prime }}{z^{ \prime  \prime }})(\frac{z^{ \prime  \prime }}{z^{ \prime }}(r -r^{ \prime }\cos (\varphi ^{ \prime })) +\frac{z^{ \prime }}{z^{ \prime  \prime }}(r -r^{ \prime  \prime }\cos (\varphi ^{ \prime  \prime }))) + \\
+\frac{z^{ \prime  \prime 2}(r -r^{ \prime }\cos (\varphi ^{ \prime })) -z^{ \prime 2}(r -r^{ \prime  \prime }\cos (\varphi ^{ \prime  \prime }))}{z^{ \prime 2} +z^{ \prime  \prime 2}}) \label{4.9}\end{multline}

with~$z^{ \prime }$,~$z^{ \prime  \prime }$ by (\ref{2.13},\ref{2.14}) and~$q0$\ given in the appendix. - As an example, the calculation is described for~$\ensuremath{\operatorname*{Re}} =30$. The parameters are~$a =300$,~$t_{e} =50000$. ~$\Delta $\ has been calculated on a horizontal cross section at~$r =10$, and two vertical sections at $x =12$ and~$x =23.5$ . Fig. 5 shows the data points for the horizontal cross section:

\begin{center}\fcolorbox[HTML]{FFFFFF}{FFFFFF}{\includegraphics[]{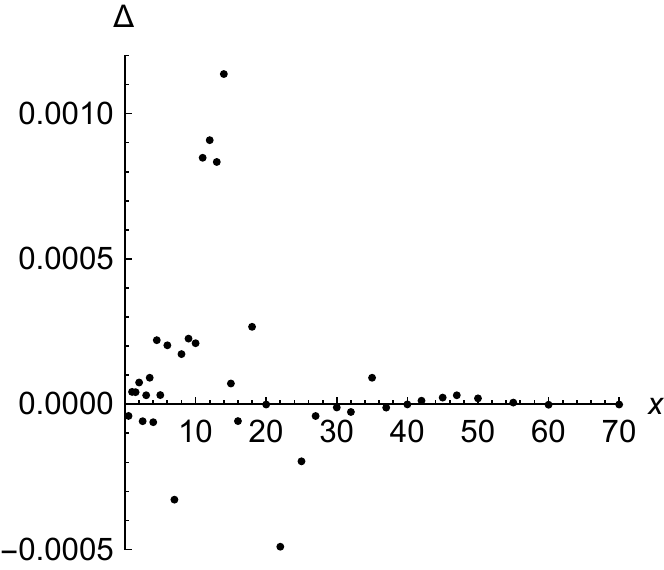}}\end{center}

\begin{center}    Fig. 5: Calculated values for~$\Delta $\ at horizontal cross section for $r =10$\end{center}\par
Immediately seen is the scatter of the points. The author presumes that the numerical calculation
errors in the neighborhood of the jet axis have not been eliminated sufficiently by the prescriptions (\ref{4.2}), (\ref{4.3}).
In order to cope with these errors, it is perhaps to be preferred to base the calculation on the Navier-Stokes equation itself instead the stream function
equation. But one should keep in mind that this is the first paper which interprets the results of \cite{pi2013};
here the aim is to have a look on the mathematical behavior of the third order term, and on the basic features of the solutions of the momentum equation.
For this purpose the evaluation to be described may be sufficient. - Moving averages (9 points each) have been calculated, and a smooth line has been drawn
to interpolate the points, as seen in Fig. 6:

\begin{center}\fcolorbox[HTML]{FFFFFF}{FFFFFF}{\includegraphics[]{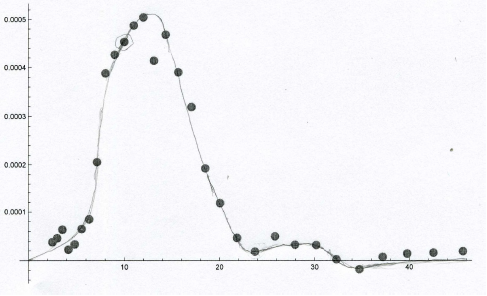}}\end{center}

\begin{center}    Fig. 6: Moving averages of fig. 5\end{center}\par
The cross point with the vertical cross section at $x =10$ has been marked. The slightly positive values for $x \geq 40$ have been ignored. The line has then been scanned with the Mathematica coordinate tool, to obtain (in this case) a set of about
50 points. In the same way, the vertical cross sections have been processed. These data are the basis for an estimate, by the Mathematica Interpolation
command, for the function $ \Delta (x ,r)$ in the range $0 \leq x ,r \leq 300$. On the 4 boundaries, $\Delta  =0$\ has been prescribed. Then the two-dimensional function has been generated with the
Interpolation command. The function behavior is somewhat ragged, but is still a well-defined set of points in the prescribed square.

\section{Approximate solution of the stream function equation}
In the
stream function equation (\ref{3.8}), on the right side we introduce, for $\Delta $, the function for laminar flow determined in the preceding section:

\begin{equation} \partial _{t}\zeta  -\frac{ \partial (\psi  ,\frac{1}{r}\zeta )}{ \partial (x ,r)} -\frac{1}{\ensuremath{\operatorname*{Re}}}\frac{1}{r}E^{2}(r\zeta ) =\Delta (x ,r) \label{5.1}
\end{equation}

\begin{center}\end{center}\par
On the right side, the Reynolds number factor is already incorporated in the function $\Delta $. (\ref{5.1}) can be seen as the first equation of an iteration procedure for calculating
the stream function. It has the same structure as the equation for laminar flow, with an additional inhomogeneous term on the right. This equation can be
solved with the NDSolve command.

The calculation has been performed, with the parameters for $\ensuremath{\operatorname*{Re}} =30$\ , in the same way as described in sec. 3. The contour plots of the solutions of
the laminar equation, and of (\ref{5.1}), look very similar. For the velocity at the axis, as a function of x, there is
practically no difference between the two calculations. But it became apparent that the method of composing the stream function field of an outer and an
inner region, as is described in sec. 3, ceases to be a practicable method in this Reynolds number range. - The calculations have been repeated for $\ensuremath{\operatorname*{Re}} =32$. For the reason mentioned, we will compare here the stream functions for the outer field only. Figs. 7 and 8 show the solutions
for laminar flow, and of eq. (\ref{5.1}):

\begin{center}\fcolorbox[HTML]{FFFFFF}{FFFFFF}{\includegraphics[]{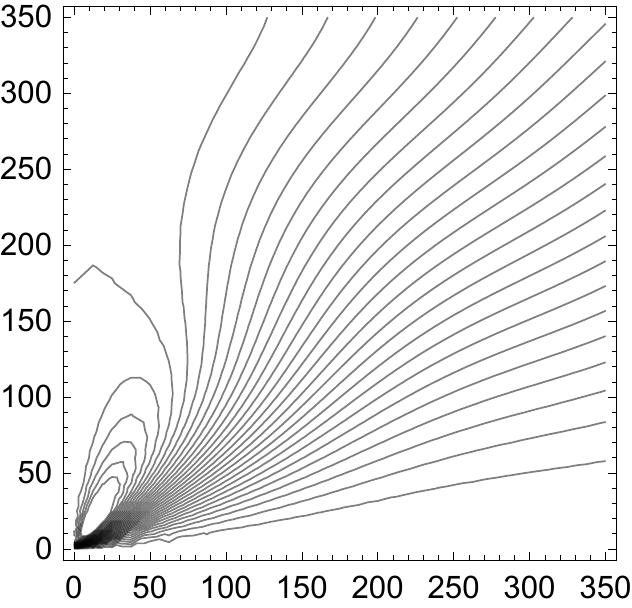}}\end{center}

\begin{center}
Fig. 7: Stream function for
$\ensuremath{\operatorname*{Re}} =32$, laminar flow\end{center}\par

\begin{center}\fcolorbox[HTML]{FFFFFF}{FFFFFF}{\includegraphics[]{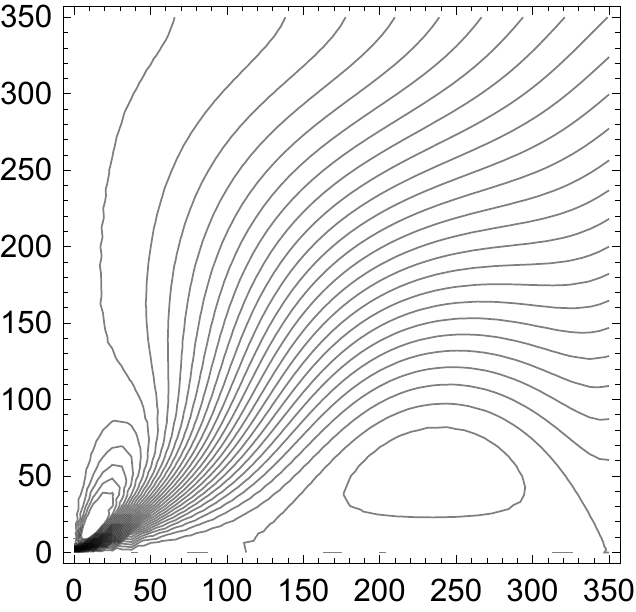}}\end{center}

\begin{center}   Fig. 8: Stream function for $\ensuremath{\operatorname*{Re}} =32$, solution of eq. (\ref{5.1})
\end{center}\par
It is seen that the third order friction term now provides a distinct distortion of the flow field
far from the nozzle. The
solution of eq. (\ref{5.1}) is the first step on an iteration which is to provide the true solution of\ eq.
(\ref{3.4}); thus, in principle it could be that the structure shown in Fig. 8 is a mathematical artefact. - On the other hand, if this result is physically relevant, it would imply that the jet is of finite length; for $x >100$, the velocity on the axis is slightly negative, i. e. the flow is oriented towards the nozzle.

\begin{center}\fcolorbox[HTML]{FFFFFF}{FFFFFF}{\includegraphics[]{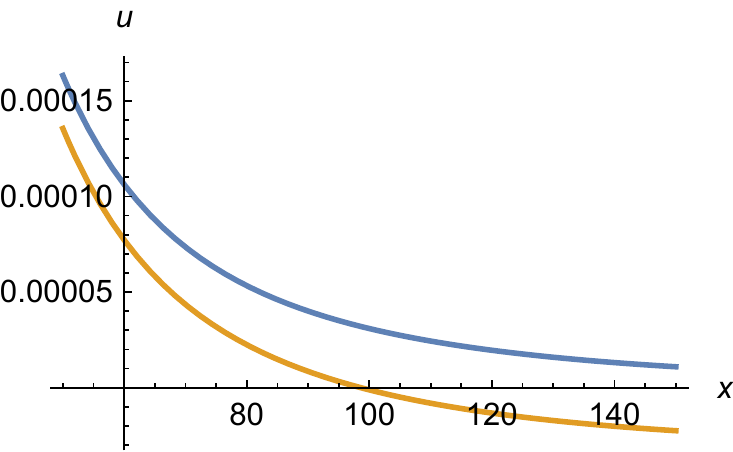}}\end{center}
\begin{center}Fig. 9: Velocity at the jet axis as a function of distance. Upper curve: solution for laminar flow\end{center}\par
Fig. 9 shows the velocity at the jet axis as a function of distance, calculated for the outer field.
Far from the nozzle the deviation from the solution for laminar flow (upper curve) is distinct. As already mentioned, for $x =100$\ the velocity becomes negative. - Finally, a calculation has been performed for $\ensuremath{\operatorname*{Re}} =35$. Here the stream function pattern shows some details which are considered to be not physically relevant. This demonstrates
that for higher Reynolds numbers it is inevitable to obtain a numerical solution of the momentum equation (or the corresponding stream function equation)
directly.

To the knowledge of the author, there are no measurements of the flow field in this Reynolds number range. Generally, such
measurements are usually performed for the axial velocity at the jet axis. The investigations by Hamel and Richter \cite{hr79}\ started
at $\ensuremath{\operatorname*{Re}} =433$. Rankin and co-workers found a formula for the axis velocity which is proportional to $\ensuremath{\operatorname*{Re}}$;\ but the actual measurements started for $\ensuremath{\operatorname*{Re}} =1000$, so it seems risky to extrapolate it for $\ensuremath{\operatorname*{Re}} =30$. - A. J. Reynolds \cite{re62}\ describes
observations of jet behavior for a range of Reynolds numbers starting at about $\ensuremath{\operatorname*{Re}} =30$. For $\ensuremath{\operatorname*{Re}} <150$, the jet is steady for a certain length, roughly proportional to $\ensuremath{\operatorname*{Re}}$, above which the jet ''breaks up''. This may be interpreted as a fluctuating jet motion for which the mean
velocity field may be calculated by solving the momentum equation described in this paper.

\section{Summary, and suggested further investigatons}
The equation for the mean velocity field in fluctuating fluid motion is applied to circular jet flow for low
Reynolds numbers. It is seen that the solutions start to deviate from the solutions of the Navier-Stokes equation when the Reynolds number exceeds $30$. Differences are most distinct in the neighborhood of the axis and in a certain distance from the orifice.

Some
additional investigations would improve the understanding of the momentum equation and its solutions:

-Find a readily available procedure
for calculating the laminar jet flow by solving the Navier-Stokes equation (or the stream function equation) which returns the boundary condition at the orifice
correctly;

-For studying the mathematical behavior of the third order term: Find the reason for and possibly eliminate
the scatter of the data;

-Since the iterative solution method employed in sec. 5 is of limited use: Develop a procedure to solve numerically
the full integro-differential equation.

In addition, some investigations must be performed before the theoretical results described
in this paper may possibly applied to problems with higher Reynolds numbers and to flow configurations with boundaries:

-Find a method
for deriving the analytical form of local equilibrium correlations. It might be possible to proceed along the lines of the papers concerning total equilibrium
correlations mentioned in\cite{pi2013} ;

-Develop
a method to find the analytical form of total equilibrium correlations for flow configurations with boundaries. \appendix

\section{Appendix: Formula for the velocity function}
The formula for the velocity function $q 0$\ (the part without a trigonometric factor) of the part $D 1$\ of the third order dissipation term has been calculated with Mathematica, brought
into TraditionalForm (partial derivatives in superscript representation), copied as a graphic and shown here as Fig. 10:

\begin{center}\fcolorbox[HTML]{FFFFFF}{FFFFFF}{\includegraphics[]{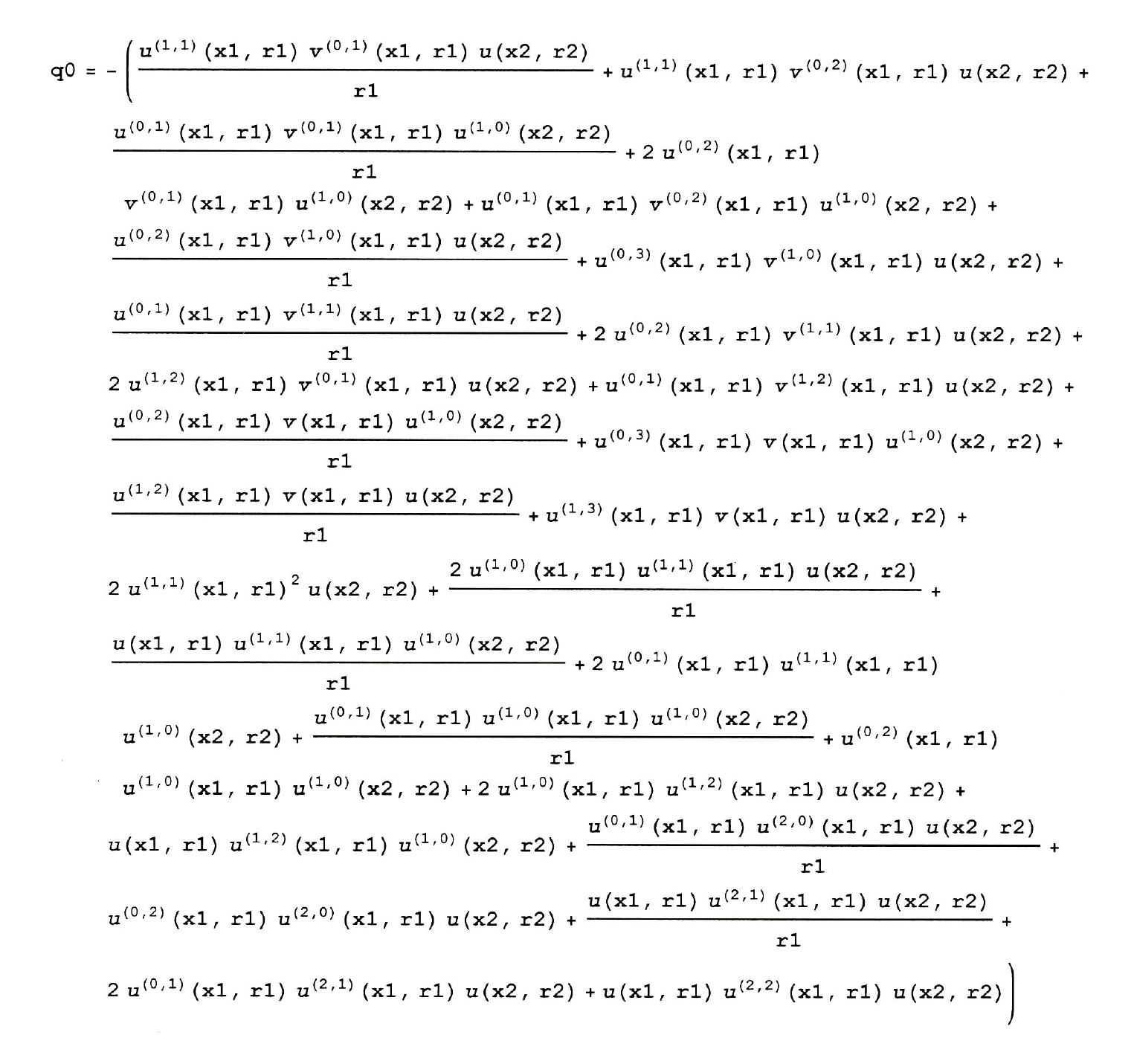}}\end{center}

\begin{center}    Fig. 10:\ Formula
for $q 0$
\end{center}\par

\end{document}